\input{psfig.sty}
\documentclass[12pt]{article}
\usepackage{graphicx}
\newfont{\feff}{cmti10}
\begin{document}

\title{Statistics of Transverse Velocity Differences in Turbulence }

\author{ Victor Yakhot\\
Department of Aerospace and Mechanical Engineering\\
Boston University, 
Boston, MA 02215 }

\maketitle

${\bf Abstract}.$
\noindent
An unusual symmetry of the equation for the generating function of transverse
velocity differences $\Delta v=v(x+r)-v(x)$ is used to obtain a closed
equation for the probability density function $P(\Delta v,r)$ in strong
three-dimensional turbulence. It is shown that 
the terms,  mixing longitudinal and
transverse components of  velocity field,  dominate the pressure
contributions.
The dissipation terms are closed on qualitative grounds. 
The resulting equation gives
the shape of the pdf,  anomalous scaling exponents and amplitudes of the moments of transverse
velocity differences.

\newpage

\noindent Intermittency of   strong turbulence seems to be a
well-established experimental 
fact. Manifested as anomalous scaling  of
velocity structure functions, this phenomenon
resisted a theoretical description for almost three decades ( for an extensive
review  see 
  [1]-[3]). Early attempts to attack the problem (cascade models) 
were based
on 
the so called refined similarity hypothesis (RSH),  connecting dissipation
fluctuations with those of the differences of longitudinal
components of velocity field. Only recently the properties of
tranverse components of velocity field  gained some interest [4]-[5].
In this paper we explore a remarkable feature of the equations for the
generating function of tranverse components of velocity field
discovered in [6] to derive the equation for the probability density (pdf) of
transverse velocity differences. The solution to  this equation describes the 
experimentally observed shape of the pdf, anomalous scaling exponents and the
amplitudes of  the
structure functions.

\noindent We are interested in statistical properties of an
incompressible  fluid
strirred by a random force in the right side of the Navier-Stokes
equations,
defined by the pair correlation function:

\begin{equation}
<f_{i}({\bf k})f_{j}({\bf k'})>\propto
P(\delta_{ij}-\frac{k_{i}k_{j}}{k^{2}})
\frac{\delta(k-k_{f})}{k^{d-1}}\delta({\bf k+k'})\delta(t-t')
\end{equation}

\noindent so that
${(f(x+r)-f(r))^{2}}\propto P(1-cos(k_{f}r))$. The integral scale of
the problem $L\approx 1/k_{f}$ and $d$ stands for  space dimensionality.
Consider two points
${\bf x}$ and ${\bf x'}$ and define ${\bf r}={\bf x-x'}$. Assuming that
the $x$-axis is parallel to the displacement vector ${\bf r}$,  one
can find 
for the separation $r$ in the 
inertial range ([1], [7], [8]) that $S_{3}=<(\Delta u)^{3}>\propto Pr$ and 
$S^{t}_{3}=\overline{(\Delta v)^{3}}
\equiv\overline{(v(x')-v(x))^{3}}= 0$ 
where $u$ and $v$ are  components of  velocity field parallel
and perpendicular to the $x$-axis  (vector ${\bf r})$, respectively. These  relations
, 
 resulting  from the Navier-Stokes equations,  are dynamic constraints on
any theory of turbulence.
The pumping power $P=O(1)$ is a constant. In what follows
we will often set $P=1$, $L=1$ 
 and restore the correct dimensionality at the end of
 calculations.
We also have [1], [7], [8]:
$r^{2-d}\partial_{r}(r^{d-1}S_{2})=(d-1)S_{2}^{t}\equiv
(d-1)\overline{(\Delta
v)^{2}}$ and for $d=3$:

\begin{equation}
6S_{3t}\equiv 6\overline{\Delta u (\Delta v)^{2}}=\partial_{r}rS_{3}
\propto Pr
\end{equation}

\noindent which will be important in what follows.

\noindent  
Following  [9], [6] and [10] we consider the $N$-point generating function:
\begin{equation}
Z=<e^{\lambda_{i}\cdot {\bf v(x_{i})}}>
\end{equation}

\noindent where the vectors ${\bf x_{i}}$ define  the positions of the points 
denoted by $1\leq i \leq N$ and summation over $x_{j}$
$x_{j}$ is assumed. 
Using the Navier-Stokes equations and incompressibility condition,
the equation for $\partial_{t} Z=\lambda_{i\mu}<(\partial_{t} v_{\mu}(x_{i}))
exp ({\bf \lambda\cdot v})>$ can be  written [6], [10]:

\begin{equation}
\frac{\partial Z}{\partial t}+\frac{\partial^{2} Z
}{\partial \lambda_{i,\mu}\partial  x_{i,\mu}}=I_{f}+I_{p}+D
\end{equation}

\noindent with the forcing, pressure and dissipation contributions
$I_{f};~ I_{p};~D$ defined below.
In what follows we will be mainly interested 
in the probability density function of the two-point velocity
differences which is obtained from (3)-(4), setting 
$\bf{\lambda_{1}+\lambda_{2}}=0$ 
(see Refs. [6], [9]-[10]), so that 
$Z=<exp{(\bf{\lambda\cdot U})}>$,  
where 
${\bf U}={\bf u(x')-u(x)}\equiv \Delta {\bf u}$.
Assuming , as in [9]-[10], that if $|U |<<u_{rms}$ and
$r<<L$, the inertial range variable $\Delta u$ is independent of the
``large-scale'' variable $U_+=u(x)+u(x+r)$ gives:

\begin{equation}
\frac{\partial Z}{\partial t}+\frac{\partial^{2} Z
}{\partial \lambda_{\mu}\partial  r_{\mu}}=I_{f}+I_{p}+D
\end{equation}

\noindent In a statistically isotropic and homogeneous flow 
the generating function can depend only on three variables:

$$\eta_{1}=r;~~ \eta_{2}=\frac{{\bf \lambda\cdot r}}{ r}\equiv 
\lambda cos(\theta);~~ \eta_{3}=\sqrt{\lambda^{2}-\eta_{2}^{2}};$$  

\noindent In these variables ($\eta_{1}=r>0$):

\begin{equation}
Z_{t}+[\partial_{\eta_{1}}\partial_{\eta_{2}}+\frac{d-1}{r}\partial_{\eta_{2}}
+\frac{\eta_{3}}{r}\partial_{\eta_{2}}\partial_{\eta_{3}}+\frac{(2-d)\eta_{2}}{r\eta_{3}}\partial_{\eta_{3}}-\frac{\eta_{2}}{r}\partial^{2}_{\eta_{3}}]Z=
I_{f}+I_{p}+D
\end{equation}

\noindent where 
\begin{equation}
I_{p}=
\lambda_{i}<(\partial_{2,i} p(2)-\partial_{1,i} p(1))e^{\bf \lambda\cdot U}>
\end{equation}
\noindent 
\begin{equation}
I_{f}=(\eta_{2}^{2}+\eta_{3}^{2})P(1-cos(k_{f}r))Z
\end{equation}

\noindent 
\begin{equation}
D=\nu\lambda_{i\mu}<(\partial^{2}_{2,j}v_{i\mu}(2)-
\partial^{2}_{1,j}v_{i\mu}(1))e^{\bf \lambda\cdot U}>=
-(\eta_{2}^{2}+\eta_{3}^{2})<({\cal E}(2)+{\cal
  E}(1))
e^{\bf \eta_{2}\Delta u +\eta_{3}\Delta v}>
\end{equation}

\noindent where, to simplify notation we set $\partial_{i,\alpha}\equiv
\frac{\partial}{\partial x.\alpha}$ and $v(i)\equiv v({\bf
  x_{i}})$ and $\lambda_{x}=\eta_{2}$ and $\lambda_{y}=\eta_{3}$. 
The dissipation rate ${\cal E}(x)=\nu (\partial_{i} v_{j})^{2}$.
The generating function, depending on a single angle $\theta$,
 can be written as:

\begin{equation}
Z=<e^{\eta_{2}\Delta u + \sqrt{d-1}\eta_{3}\Delta v}>
\end{equation}

\noindent so that any correlation function 

\begin{equation}
<(\Delta u)^{n}(\Delta v)^{m}>=(d-1)^{-\frac{m}{2}}\frac{\partial^{n}}{\partial
\eta_{2}^{n}}\frac{\partial^{m}}{\partial \eta_{3}^{m}}Z(\eta_{2}=\eta_{3}=0)
\end{equation}

\noindent Differentiating the equation (6) and setting both 
$\eta_{2}=\eta_{3}=0$ gives all known kinematic and dynamic
constraints, outlined above. It is clear from the symmetries of the problem that the
 transverse velocity difference  probability density 
is symmetric, i.e. $P(\Delta v,r)=P(-\Delta v,r)$. 
We are interested in the equation (6)-(9) in 
the limit $\eta_{2}\rightarrow 0$.
Let us first discuss some of the general properties of incompressible 
turbulence. Consider the ``one-component  forcing fucntion''
${\bf f}(x,y,z)=(f_{x}(x,y,z),0,0)$ satisfying the incompressibility
constraint
 (${\bf \nabla\cdot f}=0$).
In this case the  equation (6 ) is:

\begin{equation}
[\partial_{\eta_{1}}\partial_{\eta_{2}}+\frac{2}{r}\partial_{\eta_{2}}+
\frac{\eta_{3}}{r}\frac{\partial^{2}}{\partial_{\eta_{2}}\partial{\eta_{3}}}
-\frac{\eta_{2}}{r}\frac{\partial^{2}}{\partial \eta{_3}^{2}}-
\eta_{2}^{2}(1-cos(k_{f}r))]Z=I_{p}+D
\end{equation}

\noindent Then, setting $\eta_{2}=0$ removes all  
information about the forcing
function from the equation of motion.  The equation (12) explicitely
assumes
 that  the flow at the scale $r<<L$ is statistically isotropic and
 homogeneous
This can happen   due to  the pressure terms 
$\Delta p=-\nabla_{i}\nabla_{j}v_{i}v_{j}$, 
effectively mixing various components of the velocity field. This
universality, i.e. independence of the small-scale turbulence on the
symmetries  of
the  large-scale forcing is an assumption of this work.

\noindent  One property of  equation (12) deserves
discussion [6].  Neglecting for a time being $I_{p}$ and $D$ we have in
the limit $\eta_{2}\rightarrow 0$

\begin{equation}
[\partial_{\eta_{1}}\partial_{\eta_{2}}+\frac{2}{r}\partial_{\eta_{2}}+
\frac{\eta_{3}}{r}\frac{\partial^{2}}{\partial_{\eta_{2}}\partial{\eta_{3}}}]
Z(\eta_{2}=0,\eta_{3},r)=0
\end{equation}

\noindent and

\begin{equation}
[\partial_{r}+\frac{2}{r}+
\frac{\eta_{3}}{r}\frac{\partial}{\partial{\eta_{3}}}]Z(0,\eta_{3},r)
=\Psi(\eta_{3})
\end{equation}

\noindent where an arbitrary function $\Psi(\eta_{3})$ can be chosen
to satisfy various dynamic constraints. The fact that  equation
(14) contains only one derivative means that the underlying dynamic
equation is linear, provided both  $I_{p}$ and $D$  contributions  involve 
$\partial_{\eta_{2}}$ and the total order  of 
the original equation $n\leq 2$. It will become  clear below that this is
the case in two-dimensional turbulence, while the situation in 3D is
more complex due to  dissipation contributions,  absent in 2D.

\noindent First,  we will evaluate the pressure contribution 
$I_{p}$ which when $\eta_{2}\rightarrow 0$, can be
rewritten as:

\begin{equation}
I_{p}\approx \eta_{3}<(\partial_{y}p(0)-\partial_{y'} p(r))
exp(\sqrt{d-1}\eta_{3}\Delta v+\eta_{2}\Delta u)>
\end{equation}

\begin{equation}
\partial_{y}p(0)-\partial_{y'} p(r)=\int
k_{y}(1-e^{ik_{x}r})[\frac{k_{x}^{2}}{k^{2}}u(q)u(k-q)+\frac{k_{y}^{2}}{k^{2}}v(q)v(k-q)+\frac{k_{x}k_{y}}{k^{2}}u(q)v(k-q)]d^{2}kd^{d}q
\end{equation}

\noindent and the exponent is expressed simply as:

\begin{equation}
e^{\sqrt{d-1}\eta_{3}\Delta v +\eta_{2}\Delta u}=
exp[{\sqrt{d-1}\eta_{3}\int (1-e^{iQ_{x}r})v(Q)d^{2}Q +
\eta_{2}\int (1-e^{iQ_{x}r})u(Q)d^{2}Q}]
\end{equation}

\noindent The $k_{y}$-integration is carried out over the interval $(-\infty,
\infty)$, so that only even powers of $k_{y}$ can contribute to the
integral. The most interesting  feature of  expressions (15)-(17) is
that the additional $k_{y}$-factors can appear only from various
correlation functions resulting from the expansion of the exponents. This
is a consequence of the fact that here we are interested in the
$y$-components of the pressure gradients which are perpendicular to
the displacement vector ${\bf r}$.

\noindent Let us consider the $N=n+m$-rank  tensor 

\begin{equation}
T_{N}=<v_{i1}(x)v_{j2}(x)\cdot\cdot\cdot
v_{an}(x)v_{\alpha,n+1}(x+r)\cdot\cdot\cdot v_{\omega, n+m}(x+r)>
\end{equation}

\noindent In a statistically  isotropic flow this tensor can be
expressed only in terms of Kronekker symbols $\delta_{ij}$, 
componnets of the displacement vector $r_{\alpha}$ and some functions
$A(|{\bf r}|)$. Thus the tensor
can be represented as a sum of terms having a general structure 
$A(r)\delta_{ij}\cdot\cdot\cdot \delta_{pc}r_{j}\cdot\cdot\cdot
r_{\alpha}$. 
The product of $h$ Kronekkers contains an even ($2h$) number of
indices,  while the product of $N-2h$ components of the displacement
vector are the ones we are interested in. We have

\begin{equation}
T_{N }=\int <v_{i}(q_{1})v_{j}(q_{2})\cdot\cdot\cdot
v_{a}(q_{n})v_{\alpha}(q_{n+1})\cdot\cdot\cdot
v_{\omega}(q_{n+m})>e^{i{\bf Q\cdot r}}
\end{equation}

\noindent where ${\bf Q =q_{n+1}+q_{n+2}\cdot\cdot\cdot q_{n+m}}$. 
  The tensor $T_{N}$ can be represented as a sum of  contributions
  having the following structure:
$t_{2h}\int A(Q)\partial_{Q_{j}}\cdot\cdot\cdot
\partial_{Q_{\alpha}}e^{i{\bf Q\cdot r}}$, 
where $t_{2h}$ is a product of the $h$ Kronekker symbols.
As a consequence, a typical contribution to the correlation function
 $<(v_{i}(q_{1})v_{j}(q_{2})\cdot\cdot\cdot
v_{a}(q_{n})v_{\alpha}(q_{n+1})\cdot\cdot\cdot
v_{\omega}(q_{n+m})>$ 
can be written as:
$t_{2h}B(|Q|)Q_{j}\cdot\cdot\cdot Q_{\alpha}$ where $B(Q)\propto A(Q)/Q^{m}$. 
Now we consider each term in the right side of (16). In the limit
$\eta_{2}\rightarrow 0$ the expansion of the exponents gives only
various powers of $v(q)$'s.  The first $O(u^{2})$ term generates  
two kinds of correlation functions 

\begin{equation}
<uuv^{2n}>\propto k_{x}^{2}k_{y}^{2n}
\end{equation}

\noindent and 
$<uuv^{2n+1}>=0$. 
The second term in (16) generates a  typical contribution

\begin{equation}
<vvv^{m}>\propto k_{y}^{m}
\end{equation}

\noindent if $m$ is  even-  and is equal to zero if $m$ is an odd
number. Thus, 
both first and second terms in the right side of (16) are equal to
zero. We are left with the third contribution,   mixing the $u$ and
$v$-componnets of the velocity field.
This means that, neglecting possible infra-red divergences 
the estimate for the remaining term gives

\begin{equation}
I_{p}=b\frac{\eta_{3}}{r}<\Delta u \Delta v e^{\eta_{2}\Delta u+\eta_{3}\Delta v}>=
b\frac{\eta_{3}}{r}
\frac{\partial^{2}}{\partial \eta_{2}\partial \eta_{3}}Z(\eta_{2}=0,\eta_{3},r)
\end{equation}

\noindent This result was first obtained  for  a close- to- gaussian case
of two-dimensional flow [6].
Using some additional assumptions
about the structure of the $\Psi(\eta_{3})$ the gaussian expression
for the PDF of transverse velocity differences in two-dimensional
turbulence was obtained. Here we have proved that, in general, the only
contribution to the $I_{p}$~-~term comes from  the one mixing $u$ and $v$
components of velocity field.

\noindent The difficulty of the three-dimensional case comes from the
dissipation contributions. 
This is easily understood on the basis of the Navier-Stokes
equations. Introduce the local value of the  kinetic energy $K=v^{2}/2$ and
$Q=K(x+r)+K(r)$.
It is clear that

\begin{equation}
{\cal E}(x+r)+{\cal E}(x)\approx 
\partial_{t} Q+\partial_{2,i}v_{i}(x+r)K(x+r)+\partial_{1,i}v_{i}(x)K(x)+
\partial_{2,i}v_{i}(x+r)p(x+r)+\partial_{1,i}v_{i}(x)p(x)
\end{equation}

\noindent where the $O(\nu)$ terms are  not written down.  One can see that
the  $O(\nabla v^{3})$ dissipation contributions are much 
more complex than the
$O(\nabla v^{2})$ pressure terms, 
 derived above. Thus,

$$<({\cal E}(2)+{\cal E}(1))e^{{\bf \lambda\cdot \Delta v}}>\approx$$
$$\frac{\partial}{\partial r_{i}}<(v_{i}(2)K(2)-v_{i}(1)K(1))e^{\lambda\cdot{\bf
\Delta
v}}>-\lambda_{j}<(v_{i}(2)K(2)s_{ij}(2)-v_{i}(1)K(1)s_{ij}(1))e^{\lambda\cdot
{\bf \Delta
v}}>+R$$

\noindent where $s_{ij}(m)=\frac{\partial v_{j}(m)}{\partial x_{i}}$ and $R$
denotes the remaining contributions coming from (23) and the viscous terms. 
Assuming that the integral scale $L$
can enter the equations through the velocity differences only we make a
proposition: in the limit $\eta_{2}\rightarrow 0$:

\begin{equation}
<({\cal E}(2)+{\cal E}(1))e^{{\bf \lambda \Delta v}}>\approx
 <\Delta u\Delta v\frac{\partial \Delta v}{\partial r} e^{\lambda\cdot{\bf
 \Delta v}}>\approx 
 <\Delta u\Delta v\frac{\partial \Delta v}{\partial r} e^{\eta_{2}\Delta
 u+\sqrt{d-1}\eta_{3}\Delta v}>
\end{equation}

\noindent 
Assumption (24) is essential for the present theory.  In its spirit it 
 resembles Kolmogorov's
refined similarity hypothesis,  connecting the dissipation rate, 
averaged over a
 region of radius $r$,  with $(\Delta u)^{3}$. It will become clear
below that in the limit $\eta_{2}\rightarrow 0$ the dimensionally correct 
expression 

$$D^{o}\approx \eta^{2}_{2}\frac{\partial}{\partial r_{i}}<\Delta v_{i}(\Delta
v_{j})^{2}e^{\eta_{2}\Delta u+\sqrt{d-1}\eta_{3}\Delta v}>\propto
\eta_{2}^{2}\partial_{r}\partial_{\eta_{2}}\partial^{2}_{\eta_{3}}Z$$

\noindent cannot appear in  the equation for $Z$,
since it leads to 
exponents of the structure functions violating the Holder
inequality. It means that the model (24) can be valid only as a result of 
mutual compensation of various contributions to the expression for $D$ defined
by (9) and (23).
At the present time we cannot prove that it is so. With the assumption 
(24) we have:

\begin{equation}
D\propto \eta_{3}\partial_{\eta_{2}}\partial_{\eta_{3}}\partial_{r}Z
\end{equation}

\noindent 
Thus,  in the limit $\eta_{2}\rightarrow 0$:

\begin{equation}
[\partial_{\eta_{1}}\partial_{\eta_{2}}+\frac{2}{r}\partial_{\eta_{2}}+
\frac{\eta_{3}}{r}\frac{\partial^{2}}{\partial_{\eta_{2}}\partial{\eta_{3}}}
-\frac{\eta_{2}}{r}\frac{\partial^{2}}{\partial \eta{_3}^{2}}+ 
b\frac{\eta_{3}}{r}
\frac{\partial^{2}}{\partial \eta_{2}\partial \eta_{3}}+
c\eta_{3}\partial_{\eta_{2}}\partial_{\eta_{3}}\partial_{r}]                
Z(\eta_{2}=0,\eta_{3},r)=0
\end{equation}

\noindent where  unknown coefficients $b$ and $c$ will be determined
below. Integrating (26) over $\eta_{2}$ gives

\begin{equation}
[\partial_{\eta_{1}} +\frac{2}{r}+
\frac{\eta_{3}}{r}\frac{\partial}{\partial{\eta_{3}}}+ 
b\frac{\eta_{3}}{r}
\frac{\partial}{\partial \eta_{3}}+
c\eta_{3}\partial_{\eta_{3}}\partial_{r}]                
Z(\eta_{2}=0,\eta_{3},r)=\Psi(\eta_{3},r)
\end{equation}

\noindent We must choose $\Psi(\eta_{3})$ in such a way that the generating
function $Z(0,0,r)=1$.
Inverse Laplace transform gives the resulting equation for the pdf $P(\Delta
v,r)$

\begin{equation}
\frac{\partial P}{\partial
  r}+\frac{1+3\beta}{3r}\frac{\partial}{\partial V}VP
  -\beta\frac{\partial}{\partial V}V\frac{\partial P}{\partial r}=0
\end{equation}

\noindent Since $S_{3}^{t}=0$,  the 
coefficients in the equation for the pdf $P(V,r)=P(-V,r)$ (28) are chosen 
to give $s_{3}^{t}=\overline{|\Delta v|^{3}}=a^{3} Pr$ with an undetermined
amplitude $a^{3}$. This is an assumption of the present theory, not based  on 
rigorous theoretical considerations. 
Seeking the solution in a form $S_{n}^{t}=<(\Delta v)^{n}>\propto 
r^{\xi_{n}}$ we obtain

\begin{equation}
\xi_{n}=\frac{1+3\beta}{3(1+\beta n)}n\approx \frac{1.15}{3(1+0.05 n)}n
\end{equation}

\noindent which was derived in [10] together  with $\beta\approx 0.05$. 
The equation (29) gives:
$P(0,r)\propto r^{-\kappa}$
where $\kappa= \frac{1+3\beta}{3(1-\beta)}\approx 0.4$ for
$\beta=0.05$. Very often the experimental data are presented as $P(X,r)$ where 
$X=V/r^{\mu}$ with $2\mu=\xi_{2}\approx
0.696$ for $\beta=0.05$. This gives 

\begin{equation}
P(X=0,r)\propto r^{-\kappa +\mu}\approx r^{-0.052}
\end{equation}

\noindent The experimental data by Sreenivassan [11] give $-\kappa +\mu \approx
-0.06$. Introducing  $P(V,r)=r^{-\kappa}F(V,r)$ leads to:

\begin{equation}
(1-\beta)r\frac{\partial F}{\partial r}+\kappa V\frac{\partial
  F}{\partial V} -\beta V r\frac{\partial^{2}F}{\partial V\partial
  r}=0
\end{equation}

\noindent Next, we define $Y=V/r^{\kappa}$ so that:

\begin{equation}
(1-\beta)r\frac{\partial F}{\partial r}+2\beta\kappa Y\frac{\partial
  F}{\partial Y} +\beta\kappa Y^{2}\frac{\partial^{2}
  F}{\partial Y^{2}} 
-\beta \kappa Y r\frac{\partial^{2}F}{\partial Y\partial
  r}=0
\end{equation}

\noindent Defining $-\infty<y=Ln (Y)<\infty$, substituting this
  variable into (32) and evaluating the
  Fourier
 transform of
  the resulting 
  equation gives:

\begin{equation}
(1-\beta)r\frac{\partial F}{\partial r}+\beta\kappa
(ik-k^{2})F-ik\beta r\frac{\partial F}{\partial r}=0
\end{equation}

\noindent with the result:
$F\propto r^{\gamma(k)}$, 
where

\begin{equation}
\gamma(k)=\beta\kappa\frac{-ik+k^{2}}{1-\beta-i\beta k}Ln(r/L)
\end{equation}

\noindent with $r/L<<1$. We have to evaluate the inverse Fourier transform:

\begin{equation}
F=\int_{-\infty}^\infty dk e^{-iky}e^{\gamma(k)}
\end{equation}

\noindent in the limit $y=O(1)$ and $r\rightarrow
0$ so that $Ln(r/L)\rightarrow -\infty$. The integral can be evaluated
eaxctly. However, the resulting expression is very involved. Expanding  the
denominator in (35) gives
:

\begin{equation}
F=\int_{-\infty}^\infty dk 
e^{-ik(y+\frac{\beta\kappa |Ln (r)|}{1-\beta})}
e^{-\frac{\beta\kappa (1+\beta)|Ln (r)|}{(1-\beta)}k^{2}}
\end{equation}

\noindent and 

\begin{equation}
F\propto \frac{1}{\sqrt{\Omega (r)}}exp{(-\frac{(Ln(\xi))^{2}}{4\Omega})}
\end{equation}

\noindent with
$\xi=V/r^{\frac{\kappa}{1-\beta}}$ 
and
$\Omega (r)=4\beta\kappa\frac{1+\beta}{1-\beta}|Ln(r/L)|$.

\noindent To understand the range of validity of this expression, let
us evaluate $<V^{n}>$ using the expression (37) for the pdf. Simple 
integration, neglecting $O(\beta^{2})$ contributions, gives: 
$<V^{n}>\propto r^{\alpha_{n}}$ ~with 
$\alpha_{n}=(1+3\beta)(n-\beta (n^{2}+2))/3$. Comparing this relation
with the exact result (30) we conclude that the expression for the pdf,
calculated above, is valid in the range $n>>1$ and $\beta n<<1$. The
properties of the pdf in the range $3\leq  \xi \leq 15$ are
demonstrated on Figs. 1 and Fig.2 
 for $r/L=0.1;~0.01;~0.001$. Fig.2 shows the dependence of the pdf on $r$ for two 
values of $\xi=10; ~15$. In this range of parameter variation the
curves are reasonably well approximated by the the function 

\begin{equation}
P(\xi, r)\propto e^{-a(r)\xi^{0.9}}
\end{equation}

\noindent wwith $a(r)\approx r^{\theta}$ wwith $\theta\approx
0.12-0.15$. All quantitative predictions (29), (30) and (38)  can
  be easily veryfied exprerimentally. Indirect conformation of this result can
  be found in the experimental data by Gagne [12] on the asymmetric pdf
of longitudinal velocity differences: the curves were approximated by
an exponential with width $a(r)(r)\propto r^{\theta^{o}}$ with
  $\theta^{o}\approx 0.15$.

\noindent The log-normal distribution 
(37), derived from (26), (28),  is valid in a certain (wide but limited) 
range of $V$- variation. It is clear from (26)-(28)
that neglecting the dissipation terms ($c\propto\beta =0$) leads to 
$\xi_{n}=n/3$,  i. e. 
 disappearence  of anomalous scaling of  moments of velocity
differences.  This result agrees with the well-developed phenomenology,
attributing intermittency to the dissipation rate fluctuations: the
stronger the fluctuations, the smaller fraction of the total space
they occupy [1], [8]. To the best of our knowledge, this is the first work 
leading to multifractal distribution of velocity differences as a result of
approximations made directly on the Navier-Stokes equations. The expression
(37) is similar to the one obtained in the groundbreaking 
paper by Polyakov on the scale-invariance of strong interactions,  where the 
 multifractal scaling and the pdf were 
 analytically derived for the first time [13]-[14]. In the review paper 
[14] Polyakov noticed
that the exact result can be simply reproduced considering a cascade process 
with a heavy stream (particle) transformed into lighter streams at each step
of the cascade (fission).  
Due to   relativistic effects the higher the energy of the
particle, the smaller  the angle of a cone,  accessible to the
fragments formed as a result of  fission. Thus, the larger
 the number of a cascade step, the smaller is the
fraction of space occupied by the particles [14].  It is remarkable that the 
qualitatively similar  
picture was so successfully applied by Parisi and Frisch to the 
undertanding of  
scale invariance of strong 
turbulence [15].

\noindent To conclude: theory of turbulence needs a closure for
  the pressure gradient and dissipation terms in  the equations for the
  probability density. The task is greatly simplified by  an unusual
  symmetry of the equations for the pdf of transverse velocity differences,
  resulting in reduction of  the order of the  equation for 
the generating function $Z$. It means that, not being able to evaluate the
  entire probability density $P(\Delta u, \Delta v, r)$, we can derive an
  expression for $P(\Delta v, r)=\int_{-\infty}^{\infty}
 P(\Delta u,\Delta v, r)d\Delta u$. 
It has been shown in this work that the
  dominant contributions to the expression for the pressure gradients
  comes from the terms,  
  ``mixing''  transverse and longitudinal components of the velocity
  field. The self-interaction does not affect the  dynamics of
  transverse velocity differences.
 We believe, the expression (22) stands  on a relatively firm
  ground. The 
  closure of the 
  dissipation terms (24), (25)  is an assumption which must be tested
  experimentally. The proposed closure resembles the kolmogorov refined
  similarity hypothesis connecting the dissipation rate averaged over a
  ``ball''
of a radius $r$ around point $x$ with the third power of  velocity difference 
$(\Delta u)^{3}$. The relation (24)-(25), though,  is based on an equivalent 
 mixed
  third-order
structure  function (2),  untill now overlooked by studies of strong 
turbulence.

\newpage
\noindent $Ln(P(\xi,r))$~ ~~~~~~~~~~~~~~~~~~~~~$\xi$ \\ 
\includegraphics{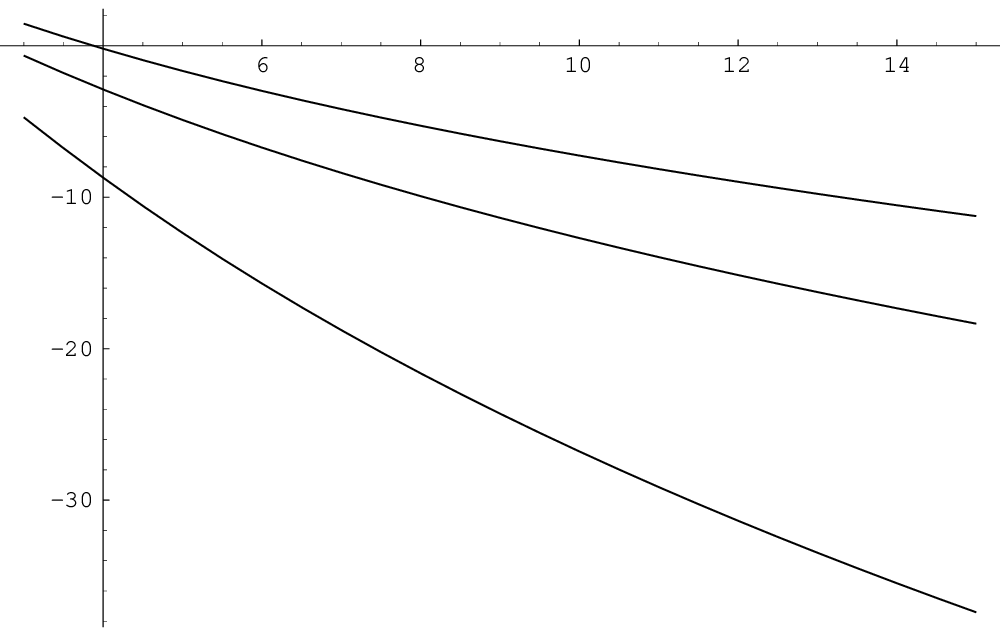}

\noindent Fig.1.~  $Ln(P(\xi,r))$.~ 
From bottom to top:  r/L=0.1;~0.01;~0.001, respectively.  
\\
\\

\noindent $ Ln(-Ln(P(\xi,r)))$\\
\includegraphics{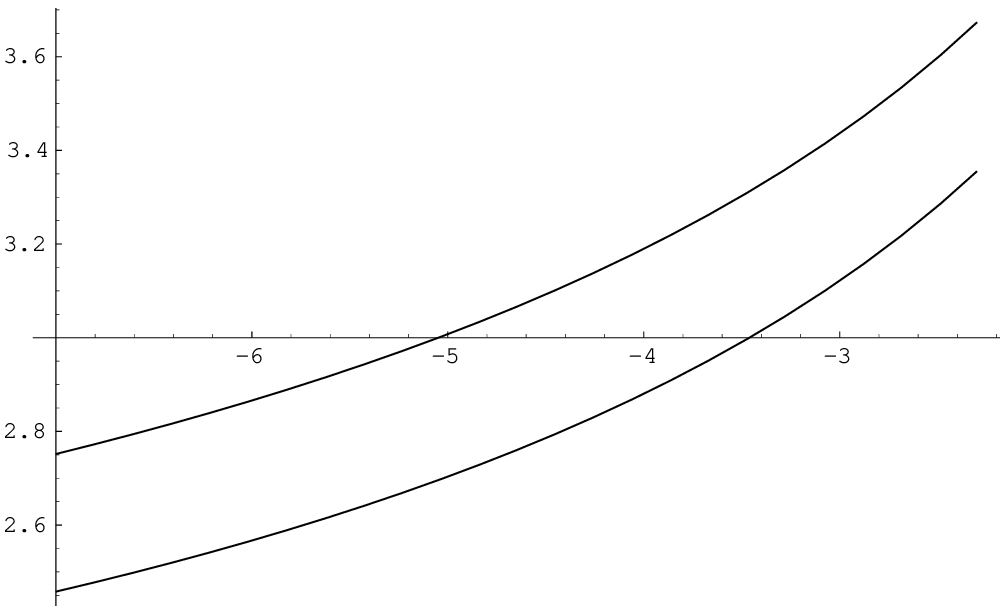}

$$Ln(r)$$
\\
\noindent Fig.2.  $ Ln(-Ln(P(\xi,r)))$ vs. $Ln(r)$ for $\xi=10$ (top) and $\xi=15$.

\noindent I would like to thank K.R. Sreenivasan for experimentally 
testing some of the
relations derived in this paper.
I am grateful to M. Nelkin, A. Polyakov and R.Tabar for helpful discussions.

\noindent {\bf references}
\\
1. U. Frisch, ``Turbulence'', Cambridge University Press, 1995
\\
2. M. Nelkin, Advances in Physics, {\bf 43}, 143 (1994)
\\
3. K.R. Sreenivasan  and R.A. Antonia, Annu.Rev.Fluid Mech. {\bf 29},
435 (1997)
\\
4. S.Y. Chen, private communication 
\\
5. A. Noullez, G. Wallace, W. Lempert, R.B. Miles and U. Frisch,
J. Fluid. Mech., {\bf 339}, 287 (1997)
\\
6. V. Yakhot, Phys.Rev.E, 1999, (in press)
\\
7.  L.D.Landau and E.M. Lifshitz, Fluid Mechanics, Pergamon Press, Oxford, 198
\\
8. A.S.Monin and A.M.Yaglom, ``Statistical Fluid Mechanics'' vol. 1, MIT Press,
Cambridge, MA (1971)
\\
9 .  A.M. Polyakov, Phys.Rev. E, {\bf 52}, 6183 (1995)
Phys.Rev. E {\bf 52}, 6183 (1995)
\\
10. V. Yakhot, Phys.Rev.E, {\bf 57}, 1737 (1998)
\\
11. K.R. Sreenivasan, private communication
\\
12. Y. Gagne, Thesis, L'Institute National Polytechnique De Grenoble, 1987.
\\
13. A.M. Polyakov, Sov. Phys. JETP {\bf 34}, 1177 (1972)
\\
14. A.M. Polyakov, ``Scale Invariance of Strong Interactions and Its
    Application to Lepton-Hadron Reactions'', preprint,
 Landau Institute for Theoretical Physics, 1971
\\
15. U. Frisch and G. Parisi, in Chaos and Statistical Mechanics, Lecture notes
     in Physics, Springer Verlag, ed U. Frisch, 1983.

\end{document}